\documentclass[12pt]{article}
\usepackage{amsmath}
\usepackage{amsfonts}
\usepackage{amssymb}
\usepackage{latexsym}
\usepackage{graphicx}
\usepackage[english]{babel}
\input epsf.sty

\begin{document}

\title{
{\normalsize \hskip4.2in USTC-ICTS-07-09} \\
{Is Noncommutative Eternal Inflation Possible?}}
 \vspace{3mm}
\author{{Yi-Fu Cai$^1$\footnote{caiyf@mail.ihep.ac.cn}, Yi Wang$^{2,3}$\footnote{wangyi@itp.ac.cn}}\\
{\small $^{1}$ Institute of High Energy Physics, Chinese Academy of
Sciences, Beijing 100049, P. R. China}\\
{\small $^{2}$ Institute of Theoretical Physics, Chinese Academy of
Sciences, Beijing 100080, P. R. China}\\
{\small $^{3}$  The Interdisciplinary Center for Theoretical Study of China (USTC), }\\
{\small Hefei, Anhui 230027, P. R. China}}
\date{}
\maketitle

\begin{abstract}
We investigate the condition for eternal inflation to take place
in the noncommutative spacetime. We find that the possibility for
eternal inflation's happening is greatly suppressed in this case.
If eternal inflation can not happen in the low energy region where
the noncommutativity is very weak (the UV region), it will never
happen along the whole inflationary history. Based on these
conclusions, we argue that an initial condition for eternal
inflation is available from the property of spacetime
noncommutativity.
\end{abstract}

\section{Introduction}

Inflation has been widely considered as a remarkably successful
theory in explaining many problems in very early universe, such as
the flatness, the horizon and the monopole problem
\cite{Guth81,Linde82,Steinhardt82,Starobinsky-inf}. It predicts
that the quantum fluctuations of the inflaton field were generated
to form today's large scale
structure\cite{Mukhanov81,Guth82,Hawking82,Starobinsky82,Bardeen83},
and these fluctuations fit with current observations of cosmic
microwave background very well\cite{CMBobserve,WMAP}.

It is a common viewpoint that eternal
inflation\cite{steinhardt-nuffield,vilenkin-eternal,linde-eternal}
can take place in a variety of inflation models. Especially, with
the monomial chaotic inflaton potential, inflation generally
becomes eternal in the high energy regime. Eternal inflation is
not only important in concept, but also provides a possible
realization for the string theory
landscape\cite{bousso-landscape,kachru-landscape,susskind-landscape,douglas-landscape}.
There have been an increasing number of papers investigating
whether we can measure eternal inflation and calculate
probabilities in the multiverse, see
references\cite{vilenkin-measure,Bousso:2006ev,Gibbons:2006pa,Linde:2006nw,Podolsky:2007vg,Li:2007rp}.

Usually, eternal inflation happens when the energy scale of the
universe is extremely high. Thus we would like to take into
consideration on more fundamental theories in logic, namely, the
string theory. In this paper, we focus on a universal property of
string theory, i.e., noncommutativity of spacetime. By considering
the effects of spacetime noncommutativity, we study the condition
for eternal inflation and its implications. As required by stringy
spacetime uncertainty
relation\cite{Li:1996rp,Douglas97,Seiberg99}, the physical time
$t_p$ and the physical length $x_p$ should satisfy
\begin{equation}
  \Delta t_p \Delta x_p \geq l_s^2~,
\end{equation}
where $l_s$ is a length scale given by the string theory.

There have been a lot of attempts to apply this uncertainty
relation into inflationary cosmology, dubbed noncommutative
inflation, see Ref. \cite{Chu-nc,Lizzi-nc,Bran2002}. More detailed
works have been investigated in a number of literature\cite{
qghuang-nc,bran-ncother,noncommutative-china,Kim-nc,Calcagni-nc,caip-nc,xue-nc}.
Here we briefly review the conception of noncommutative inflation
proposed by Brandenberger and Ho\cite{Bran2002}, and then examine
whether eternal inflation takes place with such noncommutativity.
Note that here we have slightly generalized the discussion in
\cite{Bran2002} from power law inflation into a generally
quasi-exponential inflationary scenario.

In order to introduce the noncommutativity into the 4-dimensional
Friedmann-Robertson-Walker universe, we would like to define
another time coordinate $\tau$,
\begin{equation}
  ds^2=dt^2-a^2(t)d{\vec x}^2=a^{-2}(\tau)d\tau^2-a^2(\tau)d{\vec x}^2~,
\end{equation}
where $a$ is the scale factor and we have assumed a spatially flat
universe($K=0$). For the quasi-exponential expansion where we have
imposed the usual slow roll approximation self-consistently, we
have
\begin{equation}
  d\tau=adt\simeq a_0e^{Ht}dt~, \ \ \ a\simeq H\tau~,
\end{equation}
where $H=\dot a/a$ is the Hubble parameter with the dot represents
the derivative with respect to the cosmic time $t$ and $a_0$ is an
arbitrary parameter for the scale factor at a fixed time. Then the
spacetime uncertainty relation takes the form
\begin{equation}\label{unc}
\Delta\tau\Delta x\geq l_s^2~,
\end{equation}
where $x$ is the comoving radial coordinate.
This can be realized by the commutation relation of spacetime
\begin{equation}
  [\tau,x]_*=il_s^2~,
\end{equation}
where the $*$-product is defined as
\begin{equation}
  (f*g)(x,\tau)=\left.\exp\left(-\frac{i}{2}l_s^2\left(\partial_x\partial_{\tau'}
  -\partial_\tau\partial_y\right)\right)
  f(x,\tau)g(y,\tau')\right|_{y=x,\tau'=\tau}~.
\end{equation}

Based on these considerations, for the fluctuation of the inflaton
field $\delta_q\varphi_k$, we can derive that the canonical
normalized perturbation variable $u_k\simeq a\delta_q\varphi_k$
satisfies the equation of motion
\begin{equation}
  u_k{''}+\left(k^2-\frac{z_k{''}}{z_k}\right)u_k=0~,
\end{equation}
where the prime denotes the derivative with respect to
$\tilde\eta$ defined as $d\tilde\eta\equiv a_{\rm eff}^{-2}d\tau$.
The parameter $a_{\rm eff}$ is an effective scale factor appeared
in the dispersion relation between a mode $k$ and its energy
defined with respect to $\tau$,
\begin{equation}
  a_{\rm eff}^2\equiv
  \left(\frac{\beta_k^+}{\beta_k^-}\right)^{1\over 2}=\frac{k}{E_k}~,\ \
  \beta_k^{\pm}(\tau)\equiv
  \frac{1}{2}\left[a^{\pm 2}(\tau-l_s^2k)+a^{\pm 2}(\tau+l_s^2k)\right]~,
\end{equation}
and $z_k$ is defined as
\begin{equation}
  z_k^2\equiv(\beta_k^-\beta_k^+)^{\frac{1}{2}}z^2~,\
  \ z\equiv a\frac{\dot\varphi}{H}~.
\end{equation}

As the scale factor is expanding nearly exponentially, when $l_s^2k$
is not too small compared with $\tau$ we can take the approximate
form of $\beta_k^{\pm}$ as
\begin{eqnarray}
\label{b+}
  &\beta_k^{+}\simeq{1\over 2}a^2(\tau+l_s^2k)\simeq {1\over
  2}\left[H\cdot(\tau+l_s^2k)\right]^2~,\\
\label{b-}
  &\beta_k^{-}\simeq{1\over 2}a^{-2}(\tau-l_s^2k)\simeq {1\over
  2}\left[H\cdot(\tau-l_s^2k)\right]^{-2}~.
\end{eqnarray}
Moreover, because of the relations $\Delta x\sim 1/k, \Delta
\tau\sim 1/E_k$ and the spacetime uncertainty (\ref{unc}), there
is an initial time $\tau_k$ for the perturbations to be generated,
\begin{equation}
  a_{\rm eff}(\tau)\geq a_{\rm eff}(\tau_k)=l_sk~,
\end{equation}
and the fluctuations are not allowed to exist before $\tau_k$.


Due to this initial time for the fluctuations, the quantum
fluctuations of inflaton in noncommutative environment can be
generated inside or outside the Hubble radius. These fluctuations
are called UV modes and IR modes respectively. In the UV mode
limit, the effects of spacetime noncommutativity becomes very weak
and hence the evolution of the fluctuations is similar with the
commutative case. Therefore the spectrum of UV modes is roughly
the same as in the commutative case. While in the IR region,
noncommutativity dominates the behavior of the perturbations, and
some familiar pictures will be totally modified. In this paper, we
shall use the analysis of noncommutativity mentioned above
(especially the IR region) to study eternal inflation.

This paper is organized as follows. In Section 2, we calculate the
fluctuations generated outside the horizon and provide the
constraint for eternal inflation to exist. In Section 3, we make a
conclusion and discuss some other possibilities.

\section{Fluctuations generated outside the horizon}

In this section, we investigate the quantum fluctuations generated
in noncommutative inflationary era and discuss the condition of
eternal inflation. Since in the UV case the noncommutative
property only contribute a few corrections on the usual
perturbation theory\cite{Bran2002,qghuang-nc}, the condition for
the eternal inflation to happen is basically the same as the
commutative one. Therefore, we focus our consideration on the IR
case that the quantum fluctuations are generated outside the
horizon. At the time the fluctuations start to be generated, the
effective scale factor is
\begin{equation}
  a_{\rm eff}(\tau_k)=l_s k.
\end{equation}
Making use of Eq. (\ref{b+}) and (\ref{b-}), and the fact that the
Hubble scale is larger than the noncommutativity scale in this
case, we get the initial time $\tau_k$ and the initial scale
factor $a_k$ as
\begin{equation}\label{tausol}
  \tau_k=\sqrt{l_s^4k^2\left(1+{1\over l_s^2H^2}\right)}\simeq
  l_s^2k, \ \ \ a_k\simeq Hl_s^2k
\end{equation}

Since the IR modes are generated outside the horizon, it is
required that $k<a_kH$. By using the second relation in Eq.
(\ref{tausol}), we can see that
\begin{eqnarray}\label{ircon}
H>l_s^{-1}~
\end{eqnarray}
in the IR case. This verifies the physical picture that $H$ should
be larger than the noncommutativity scale.

Now we calculate the quantum fluctuation in the momentum space
$\delta_q\varphi_k$. $\delta_q\varphi_k$ is linked to the
canonical perturbation $u_k$ by $u_k\simeq a\delta_q\varphi_k$,
and when the perturbation begins to be generated the initial $u_k$
was canonically normalized as $u_k\simeq \frac{1}{\sqrt{2k}}$.
Consequently, when $\delta_q\varphi_k$ is born its amplitude can
be given by
\begin{equation}
  \delta_q\varphi_k\simeq\frac{1}{\sqrt{2k}}\frac{1}{Hl_s^2k}~.
\end{equation}
After that, the fluctuations outside the horizon are nearly
frozen. By transforming to the coordinate space, we obtain the
relation
\begin{eqnarray}
\left<\delta_q\varphi^2\right>=\int_{k=aH}^{k=(e\times a)H}
\frac{dk}{k}\frac{k^3}{2\pi^2}\delta_q\varphi_k\delta_q\varphi_{-k}\simeq\left(\frac{1}{2\pi}\frac{1}{Hl_s^2}\right)^2
\end{eqnarray}
during one Hubble time, and correspondingly the IR quantum
fluctuation $\delta_q\varphi$ per e-fold in the noncommutative
spacetime is given by
\begin{equation}
  \delta_q\varphi\simeq\frac{1}{2\pi}\frac{1}{Hl_s^2}~.
\end{equation}
Note that this result is strongly different from the commutative
one $\delta_q\varphi\simeq H/2\pi$. Due to this, the physics of
eternal inflation is greatly modified in the noncommutative case
and the condition for eternal inflation to happen needs to be
reconsidered.

As usual, the classical motion of the inflaton during one Hubble
time takes the form
\begin{equation}
  \delta_c\varphi \simeq \dot\varphi H^{-1} \sim {V_{\varphi}\over H^2}~,
\end{equation}
where $V_{\varphi}$ denotes $dV(\varphi)/d\varphi$. And as usual,
the condition that inflation becomes eternal is roughly
$\delta_q\varphi>\delta_c\varphi$, or
\begin{equation}\label{main}
  H>V_{\varphi}l_s^2~.
\end{equation}
Compared with the eternal inflation condition in the commutative
case $H^3>V_{\varphi}$, the noncommutative eternal inflation is
more unlikely to happen. It is because the quantum fluctuation is
generally smaller due to the suppression by the Hubble parameter.
In the following, we shall consider two explicit examples with the
chaotic potentials.

Firstly, let us consider the model $V=\frac{1}{2}m^2\varphi^2$.
From the condition (\ref{ircon}) we can see that the inflation is
in the IR region when $\varphi>\frac{M_p}{l_s m}$ while in the UV
region when $\varphi<\frac{M_p}{l_s m}$.
Consequently, the condition that inflation has become eternal in
the UV region is
\begin{equation}\label{m2phi2pre}
  \varphi_{\rm IR}\equiv\frac{M_p}{l_sm}>\sqrt{\frac{M_p^3}{m}}, \ \
  \ m<{1\over M_p l_s^2}~.
\end{equation}
Interestingly, we note that from (\ref{main}) the eternal
inflation will nicely continue in the IR region if
\begin{equation}\label{m2phi2}
  m<{1\over M_pl_s^2}.
\end{equation}
To substitute the well-known value $m \sim 10^{-6}
M_p$\cite{Linde:2007fr} into (\ref{m2phi2}), we obtain the result
that eternal inflation is allowed to happen only if $M_s \simeq
l_s^{-1} > 10^{-3}M_p$.

So we conclude that if inflation has become eternal before
entering the IR region, it can continue to be eternal in the IR
region. But the amplitude of fluctuation can not grow to larger
values because $\delta_q\varphi$ is now suppressed by a factor of
$1/H$. This is greatly different from the commutative eternal
inflation and may constrain the initial condition space for the
eternal inflation. On the other hand, if inflation has not become
eternal before entering the IR region, then the inequality
(\ref{m2phi2}) does not hold. In this case inflation will never
become eternal because the quantum fluctuation can not be large
enough.

\begin{figure}
\begin{center}
\includegraphics[width=3.7in]{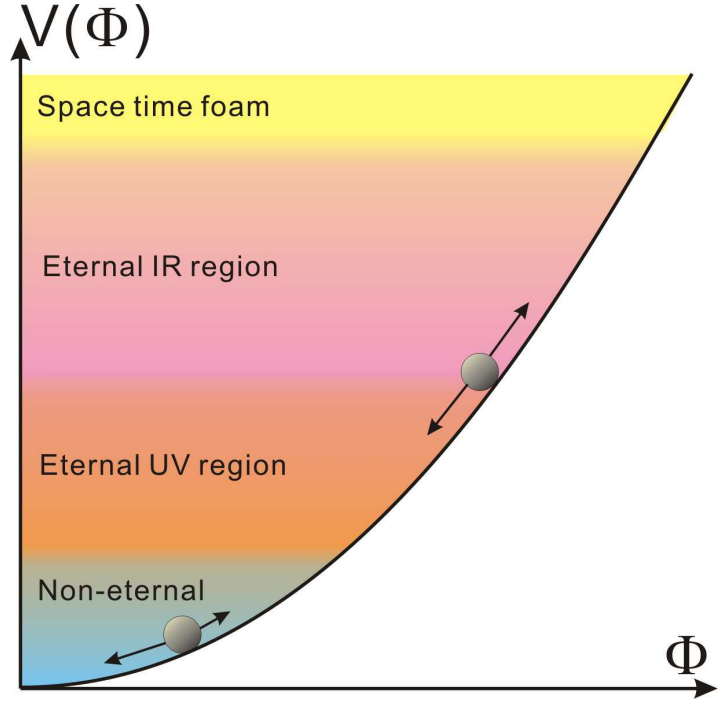}
\end{center}
\caption{Evolution of the chaotic inflation field with
$V(\varphi)=\frac{1}{2}m^2\varphi^2$ in the noncommutative
spacetime. Here we have assumed $M_s\simeq l_s^{-1}$ to be large
enough so that eternal inflation can take place in the UV region.
In this case, inflation is eternal in part of the noncommutative
UV region and the whole IR region.} \label{fig:primordial2}
\end{figure}

As a second example, consider $V=\lambda M_p^{4-p}\varphi^p$ with
$p>2$. Then similarly, the condition for entering the IR region
$H>1/l_s$ requires
\begin{equation}\label{p1}
  \varphi>\lambda^{-\frac{1}{p}}l_s^{-\frac{2}{p}}M_p^{\frac{p-2}{p}}.
\end{equation}
From the inequality (\ref{main}), we obtain the condition for
inflation to be eternal in the IR region as follow,
\begin{equation}\label{p2}
  \varphi< p^{-\frac{2}{p-2}} \lambda^{-\frac{1}{p-2}}l_s^{-\frac{4}{p-2}}M_p^{\frac{p-6}{p-2}}
\end{equation}
In order that (\ref{p1}) and (\ref{p2}) has overlap and inflation
can be eternal in the IR region, we need
\begin{equation}\label{pcond}
\lambda< p^{-p} \left(\frac{1}{l_sM_p}\right)^{p+2}~.
\end{equation}
If (\ref{pcond}) is satisfied, the noncommutative eternal
inflation is allowed to exist in the IR region, and there is an
upper bound for the energy density of eternal inflation which
arises from (\ref{p2}). Since the inflaton field can not climb
higher than the bound (\ref{p2}) with large probability, this
provides a possible initial condition for eternal inflation.

Now let us examine whether there is eternal inflation at all in
the $p>2$ model above. For the eternal inflation to take place in
the UV region, we again obtain the relation
\begin{equation}
  \lambda< p^{-p} \left(\frac{1}{l_sM_p}\right)^{p+2}~,
\end{equation}
which is the same as the IR region eternal condition (\ref{pcond}).

So we get the similar conclusion with the
$\frac{1}{2}m^2\varphi^2\ (p=2)$ model that if eternal inflation
takes place in the UV region, it is allowed to extend into the IR
region. On the other hand, if inflation can not be eternal in the
UV region, unfortunately there will not be any eternal inflation
in the whole noncommutative inflationary history. Besides, there
is an interesting difference between $p>2$ and $p=2$ cases that,
in the $p>2$ case noncommutative eternal inflation requires an
upper bound on the inflaton while this bound does not exist in the
$p=2$ case.

For example, in the $\lambda\varphi^4$ model, if we expect eternal
inflation to take place, then we need the noncommutative scale
$l_s^{-1}>\lambda^{\frac{1}{6}}M_p$. To apply the data $\lambda
\sim 10^{-14}$, we obtain that this scale is around $10^{-2}M_p$.
If $l_s^{-1}$ is just above this scale, the upper bound (\ref{p2})
on $\varphi$ is around $10^3 M_p$ and the corresponding energy
density is about $10^{-2} M_p^4$. Consequently, we conclude that
the limitation from noncommutativity can be much stronger than
that from the Planck density. Such a string scale can be realized
easily if we do certain compactification on a large manifold,
which should be common and have a large prior probability in the
string landscape.

\begin{figure}
\begin{center}
\includegraphics[width=3.6in]{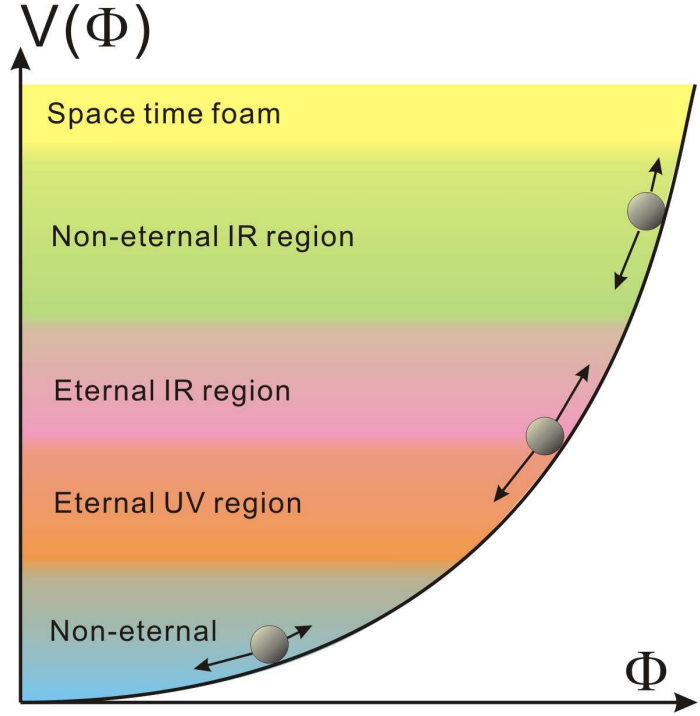}
\end{center}
\caption{Evolution of the chaotic inflation field with
$V(\varphi)=\lambda\varphi^4$ in the noncommutative spacetime.
Here we have assumed $M_s\simeq l_s^{-1}$ to be large enough so
that eternal inflation can take place in the UV region. In this
case, inflation is eternal in part of the noncommutative UV region
and part of the IR region. This figure is different from Fig.
\ref{fig:primordial2} that eternal inflation can not happen in the
green part of IR region where the energy scale is extremely high.
} \label{fig:primordial4}
\end{figure}

On the other hand, such a scale $l_s^{-1}\sim 10^{-2}M_p$ or
$l_s^{-1}\sim 10^{-3}M_p$ is considerably higher than the scale
for the final 60 e-folds inflation. So if we can observe the
noncommutativity in future CMB experiments\cite{caip-nc}, then
eternal inflation is not expected to have happened.

\section{Conclusion and discussions}

Spacetime noncommutativity, predicted by string theory, has become
a fundamental principle and been studied in a number of literature
(see e.g. \cite{Bigatti00,Alekseev00,Chu02}). This principle
brings new physics when applied into inflation theory. In this
paper,
we have seen that for the chaotic inflationary potential, the
scenario of eternal inflation in noncommutative spacetime is
remarkably different from the usual one.
If inflation do not become eternal in the UV region, then it can
never have happened.

We have also discussed that if eternal inflation happens
both in the UV region and the IR region, an initial condition for
eternal inflation can be provided or constrained. From the
derivation in this paper, we can see that $\delta_q\varphi$
becomes smaller and smaller along the potential when eternal
inflation enters IR region.
Therefore, for the $p=2$ model the initial condition space is
greatly reduced; while for the $p>2$ model there is an upper bound
for $\varphi$ explicitly. Eternal inflation can not climb into
higher energy regions than this bound. This provides an initial
condition for eternal inflation. As is discussed in
\cite{Bousso:2006ev, Podolsky:2007vg}, initial conditions may be
essential for predictions in the multiverse. The initial condition
discussed in this paper provides a possible solution for the
initial condition problem and can be used in calculating the
eternal inflationary probabilities.

In the derivation made in Section 2, we have used the standard
method according to a number of calculations for noncommutative and
non-eternal inflation in the literature. However, generally there
are several possibilities which may change the results we obtain in
this paper.

The simplest possibility is that there is no spacetime
noncommutativity at all or the scale of noncommutativity is of the
same order as the Planck scale. Therefore, the noncommutativity
will not alter the picture of the usual eternal inflation.

As another possibility, the noncommutative field theory may be not
precise enough to describe the generation of the quantum
fluctuations. This saturation is somewhat similar to that
inflationary fluctuations described by the common quantum field
theory suffers from the transPlanckian
problems\cite{Martin:2000xs}. To see this problem in the
noncommutative field theory, we note that we have used the
relation $\Delta x\sim 1/k$ in the calculations of noncommutative
inflation. However, this relation does not hold in theories with
certain UV-IR relations while UV-IR relations arise commonly
together with noncommutativity. So there is possibility that the
perturbations are generated even if $k>a_{\rm eff}/l_s$, but it is
still an open issue for us to fully understand the physics in this
region.

Finally, the background geometry may be affected considerably by
noncommutativity in the IR region. However, up to now this case
has not been carefully studied even in the non-eternal
inflationary regime.

\section*{Acknowledgments}

We thank Bin Chen, Miao Li, Yun-Song Piao and Xinmin Zhang for
helpful discussions. This work is supported in part by National
Natural Science Foundation of China under Grant Nos. 90303004,
10533010, 19925523 and 10405029, and by the Chinese Academy of
Science under Grant No. KJCX3-SYW-N2. The author Y.W.
acknowledgments grants of NSFC.


\end{document}